\documentclass[conference,a4paper]{IEEEtran}
\usepackage{xcolor}
\usepackage{balance}
\usepackage{booktabs}
\usepackage{multirow}
\usepackage{gensymb}
\usepackage{array}

\ifCLASSINFOpdf
  \usepackage[pdftex]{graphicx}
\else
  \usepackage[dvips]{graphicx}
\fi
\ifCLASSOPTIONcompsoc
 \usepackage[caption=false,font=normalsize,labelfont=sf,textfont=sf]{subfig}
\else
 \usepackage[caption=false,font=footnotesize]{subfig}
\fi
\begin{document}
\title{A Wideband Tri-Band Shared-Aperture Antenna Array for 5G and 6G Applications}

\author{\IEEEauthorblockN{
Shang-Yi Sun\IEEEauthorrefmark{1},   % 1st author, 1st affiliations
Can Ding\IEEEauthorrefmark{1},   % 2nd author, 2nd affiliations
Hai-Han Sun\IEEEauthorrefmark{2},    % 3rd author, 3rd affiliations
Alessio Monti\IEEEauthorrefmark{4},
Y. Jay Guo\IEEEauthorrefmark{1}      % 4th author, 4th affiliations
}                                     % ...
%\\
\IEEEauthorblockA{\IEEEauthorrefmark{1}% 1st affiliations
Global Big Data Technologies Centre, University of Technology Sydney, Sydney, Australia, Can.Ding@uts.edu.au}
\IEEEauthorblockA{\IEEEauthorrefmark{2}% 2nd affiliations
Department of Electrical and Computer Engineering, University of Wisconsin-Madison, Madison, USA}
\IEEEauthorblockA{\IEEEauthorrefmark{4}
Department of Industrial, Electronic and Mechanical Engineering, Roma Tre University, Rome, Italy}
}



\maketitle

\begin{abstract}
This work presents a wideband tri-band shared-aperture antenna array covering the 5G mid-band and 6G centimetric band. The challenge of scattering and coupling suppression is holistically addressed across the wide bands. Guided by characteristic mode analysis (CMA), a segmented spiral radiator is developed to mitigate high-frequency scattering and coupling while maintaining low-frequency radiation performance. Compared with a conventional tube radiator, the proposed spiral achieves a reduced radar cross-section (RCS) over 4.7–21.5 GHz (128.2\%). With the aid of serial resonators, the segmented-spiral dipole achieves impedance matching in the low band (LB, 3.05–4.68 GHz, 42.2\%), covering the 5G band (3.3–4.2 GHz), while additional suppressors further reduce cross-band coupling. The middle band (MB) and high band (HB) antennas operate at 6.2–10.0 GHz (46.9\%) and 10.0–15.6 GHz (43.8\%), respectively, collectively covering the anticipated 5G-Advanced and 6G bands (6.425–15.35 GHz). Both the MB and HB antennas employ a planar magnetoelectric (ME) dipole structure to avoid common-mode resonances within the LB and MB and to minimize cross-band scattering in the HB. The proposed array maintains undistorted radiation patterns and better than 20 dB port isolation between any two ports across all three bands.
\end{abstract}

\vskip0.5\baselineskip

% \fi
% \IEEEpeerreviewmaketitle

\section{Introduction}

The rapid evolution of the mobile communication technologies, coupled with the continuous pursuit of cost-efficiency and miniaturization, requires antennas operating at different frequencies to share an extremely limited space to simultaneously support various standards. The co-existence of different antennas results in scattering and coupling, which leads to severe distortion in radiation patterns, and degradation of isolation and impedance matching \cite{Book}. Wideband suppression of the scattering and coupling remains a critical challenge in designing high-performance multi-band antenna arrays.

In dual-band arrays, cross-band scattering occurs between the low-band (LB) and high-band (HB) antennas. To suppress HB scattering from LB antennas, techniques such as cloaks, slots, chokes, or frequency selective surfaces (FSS) \cite{S1, S8, S10} are co-designed with LB antennas to generate reversed currents or block induced currents. The LB scattering, caused by the common-mode resonance of HB antennas, can be mitigated by connecting an inductor or capacitor in series with the HB balun to shift the resonance out of the LB \cite{S2}. Additionally, filtering techniques such as parasitic loops and branches are commonly used to reduce cross-band coupling.

For tri-band arrays, suppression of scattering and coupling is more challenging than in dual-band arrays, as interference arises between any two of the three antenna types. Recent studies mainly adapt dual-band suppression techniques to tri-band designs. To reduce scattering in both the middle band (MB) and HB, dual-passband FSS structures are integrated into the LB radiators \cite{T5}. Chokes, slots, and series inductors are simultaneously used in \cite{T1}, while slots and FSS-based MB radiators are jointly employed in \cite{T3} to mitigate various forms of scattering. In \cite{T2}, a stacked configuration and FSS-based MB radiators are adopted to suppress scattering and coupling. Nevertheless, due to the increased challenges, existing tri-band arrays still exhibit relatively narrow bandwidths.

6G has entered the research stage and is expected to utilize multiple bands within 7.125-15.35 GHz to achieve a balance between capacity and coverage \cite{6G1}, \cite{6G3}, \cite{6G2}. The 6.425–7.125 GHz, primarily allocated for 5G-Advanced, may also serve as a foundational band for 6G. Higher and additional bands introduce new challenges for multi-band array design.

\begin{figure}[!t]
\centering
\includegraphics[trim=0 10 0 0,scale=1.08]{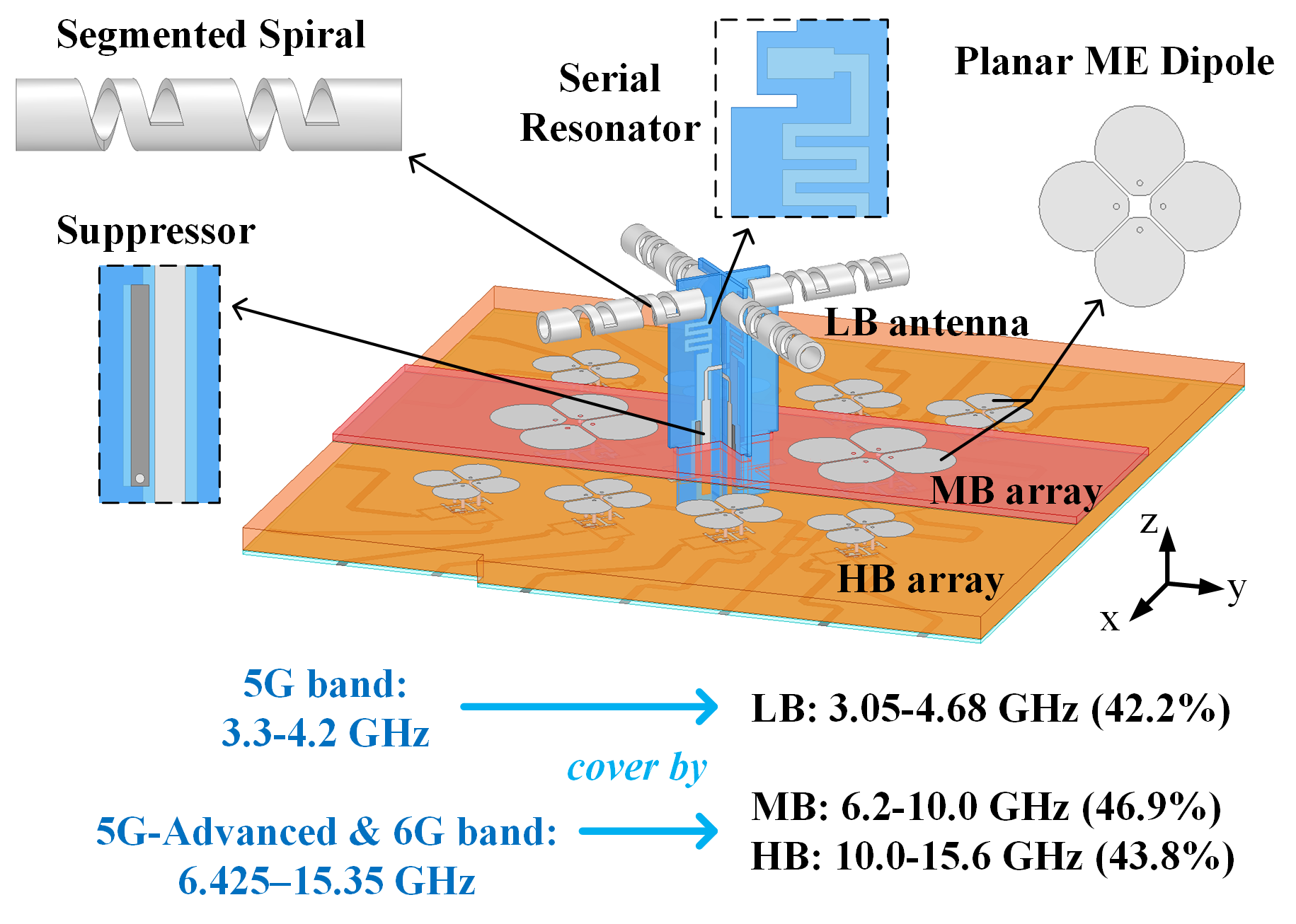}
\caption{Configuration of the developed tri-band array.}
\label{Tri_Band_Array}
\vspace{-8pt}
\end{figure}

In this work, a tri-band 5G/6G antenna array, as shown in Fig. \ref{Tri_Band_Array}, is developed with operation bandwidths in the LB, MB, and HB of 3.05-4.68 GHz (42.2\%), 6.2-10.0 GHz (46.9\%), and 10.0-15.6 GHz (43.8\%), respectively. Across all three bands, the radiation pattern distortions caused by scattering have been effectively mitigated or avoided, and all cross-/in-band coupling has been suppressed to below -20 dB.

\section{Design of LB, MB, and HB Antennas}

The cross-band scattering, cross-band coupling, and in-band coupling in the MB/HB originate from MB/HB-induced currents on the LB radiators \cite{Book}. According to characteristic mode theory, the total induced current on the LB radiator under MB/HB excitation can be approximated by a combination of a few dominant modes with large modal weighting coefficients (MWCs). Effective suppression of scattering and coupling can therefore be achieved by reducing the $|$MWC$|$ values of these significant modes \cite {CMA1}.

\begin{figure}[!t]
\centering
\includegraphics[trim=0 10 0 0,scale=1]{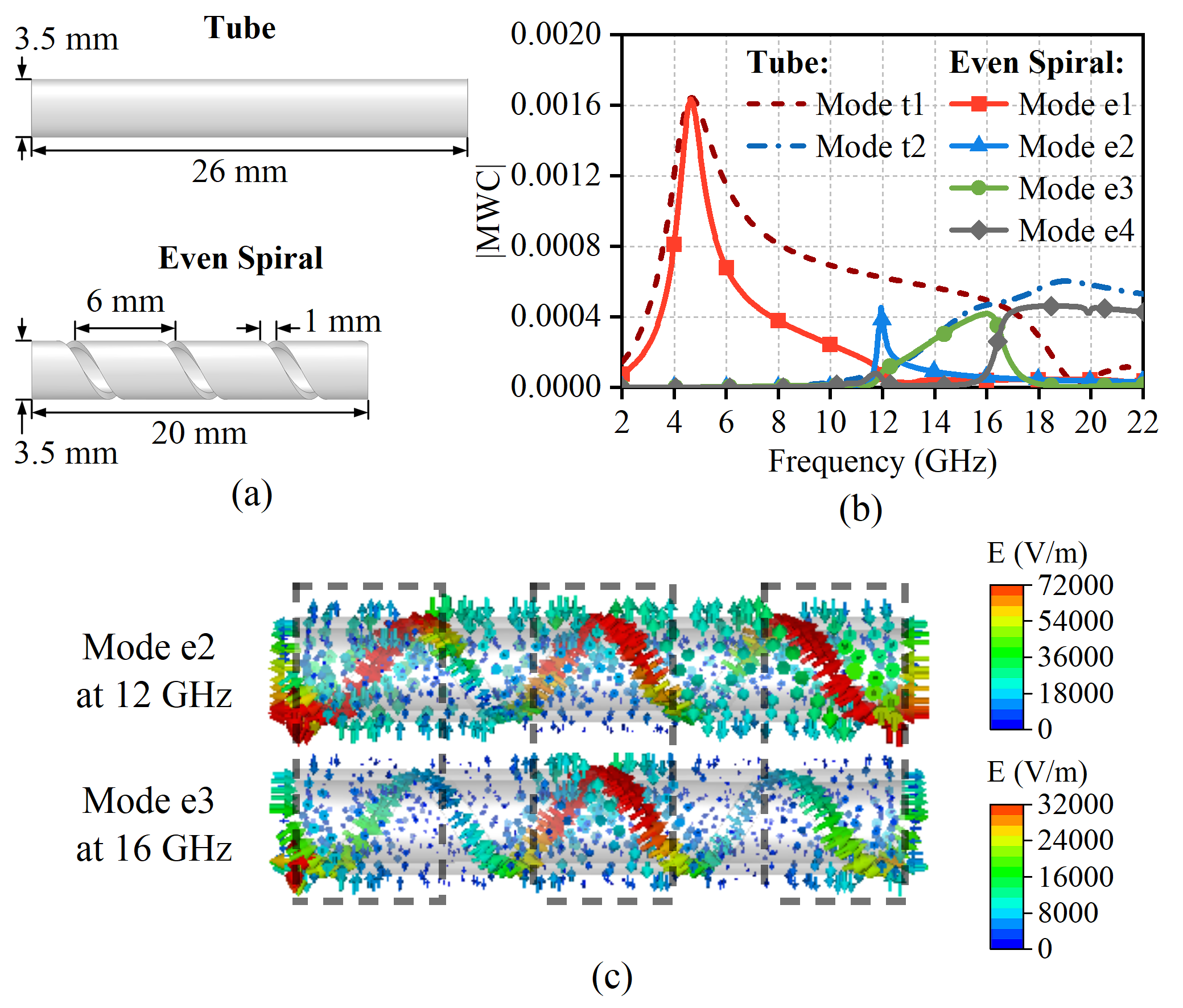}
\caption{(a) Geometry of the tube and even spiral, (b) $|$MWC$|$ of the tube and even spiral, and (c) modal E-field distribution of Mode e2 and Mode e3.}
\label{CMA_Even}
\end{figure}

The cylindrical tube in Fig. \ref{CMA_Even}(a) serves as reference case in the following sections. As shown in Fig. \ref{CMA_Even}(b), Mode t1 is its dominant mode, with $|$MWC$|$ values remaining high at higher frequencies, indicating strong excitation by high-frequency waves. This work aims to broaden the scattering and coupling suppression bandwidth while maintaining the LB antenna’s matching performance. Hence, it is crucial to reduce the $|$MWC$|$ of Mode t1 at higher frequencies while keeping that at lower frequencies undegraded.

\begin{figure}[!t]
\centering
\includegraphics[trim=0 10 0 0,scale=1]{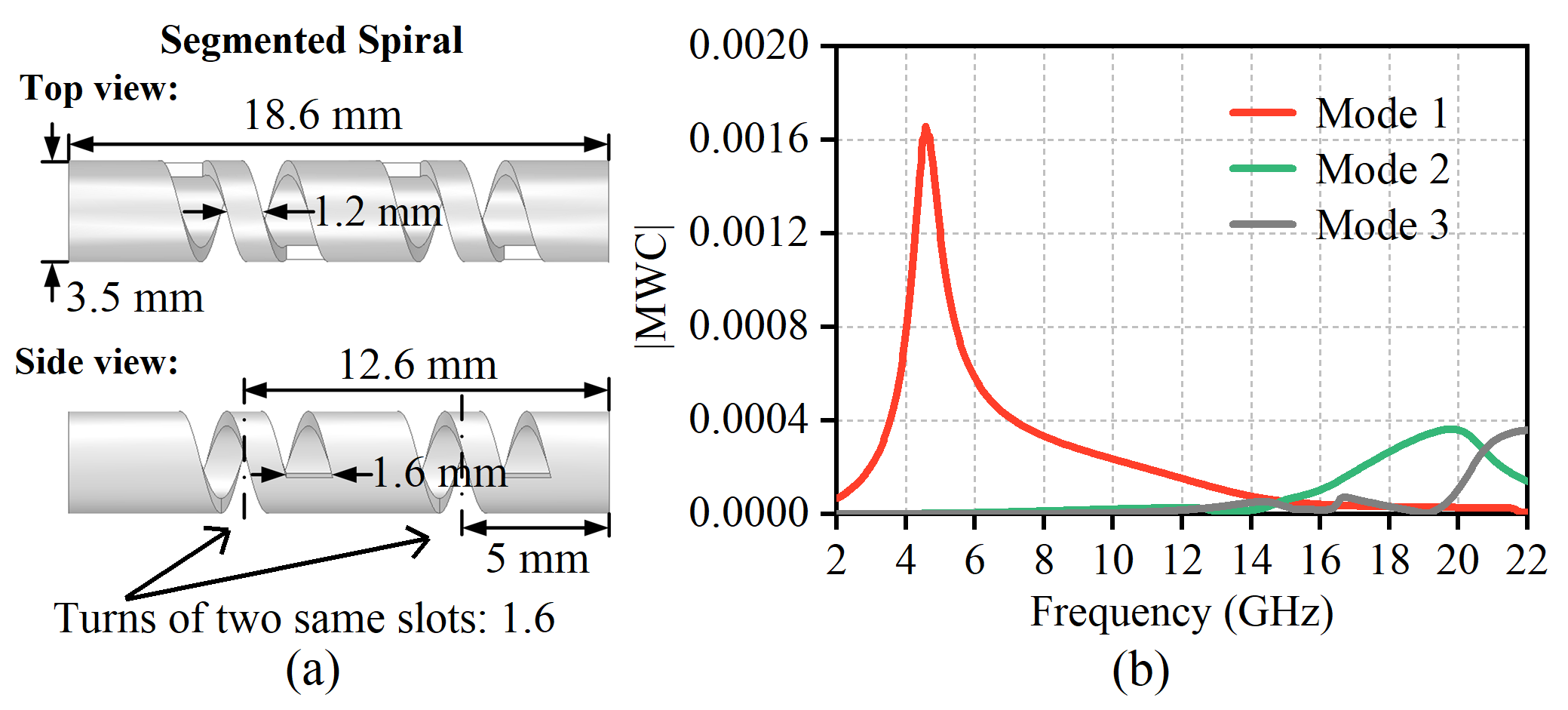}
\caption{(a) Geometry and (b) $|$MWC$|$ of the segmented spiral.}
\label{CMA_Uneven}
\end{figure}

As shown in Fig. \ref{CMA_Even}(a), the even spiral developed in \cite{S9} offers a promising solution. Fig. \ref{CMA_Even}(b) shows the $|$MWC$|$ values of significant modes in the even spiral, where Mode e1 corresponds to Mode t1 of the tube and is effectively suppressed at higher frequencies. However, additional modes e2, e3, and e4 appear, limiting the suppression bandwidth. Therefore, Modes e2 and e3 should be eliminated or shifted to higher frequencies. These modes can be suppressed by short-circuiting the slot areas with strong E-fields, as indicated by the dashed lines in Fig. \ref{CMA_Even}(c).

\begin{figure}[!t]
\centering
\includegraphics[trim=0 10 0 0,scale=1]{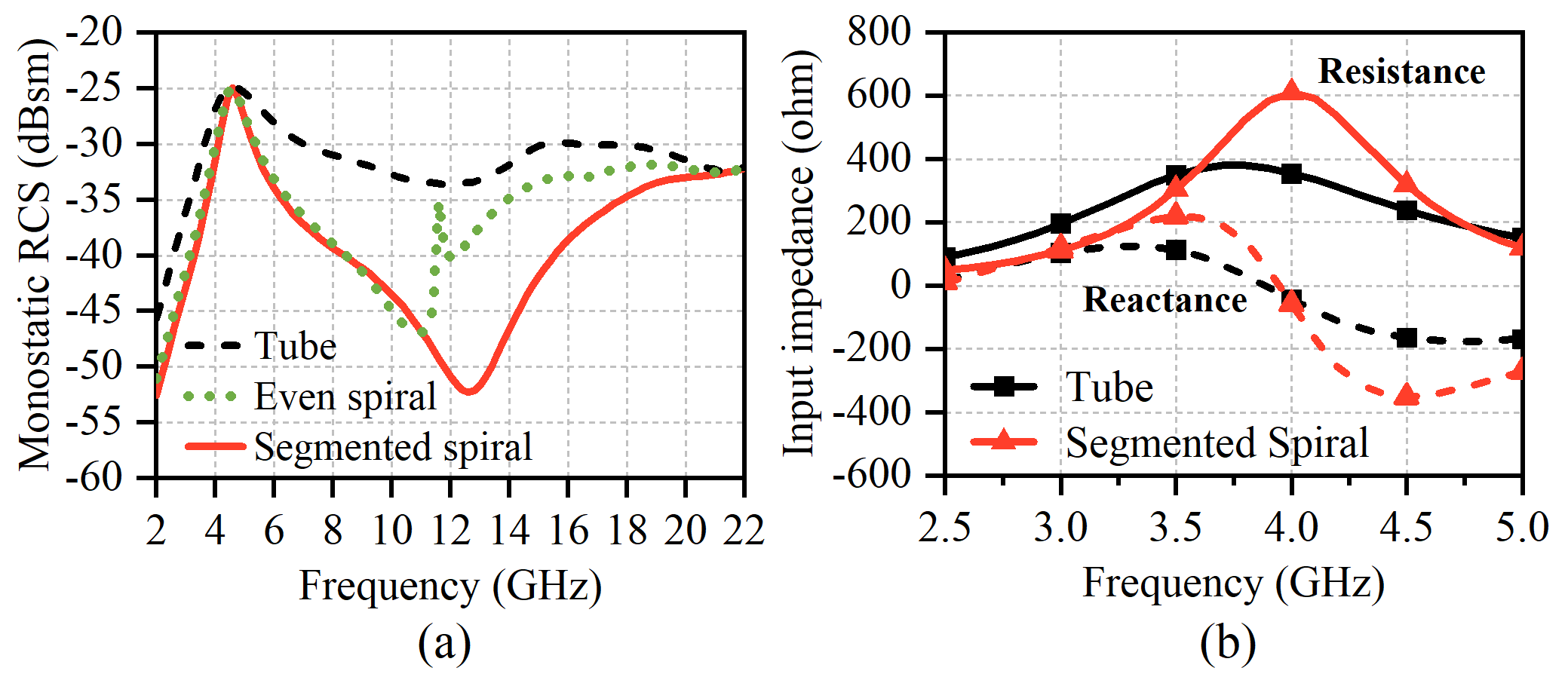}
\caption{(a) Monostatic RCSs of the tube, even spiral, and segmented spiral. (b) Input impedance of the LB dipole using the tube or segmented spiral.}
\label{RCS}
\end{figure}

Short-circuiting the even spiral forms a segmented spiral, as shown in Fig. \ref{CMA_Uneven}(a). As given in Fig. \ref{CMA_Uneven}(b), Modes e2 and e3 in Fig. \ref{CMA_Even}(b) are completely eliminated. To demonstrate the superiority of the segmented spiral in scattering suppression, its monostatic radar cross sections (RCSs) are compared with the reference tube and even spiral in Fig. \ref{RCS}(a). The even spiral is capable of reducing the RCS, but only over a limited frequency range, whereas the segmented spiral achieves wideband suppression from 4.7 to 21.5 GHz (128.2\%).

\begin{figure}[!t]
\centering
\includegraphics[trim=0 10 0 0,scale=0.9]{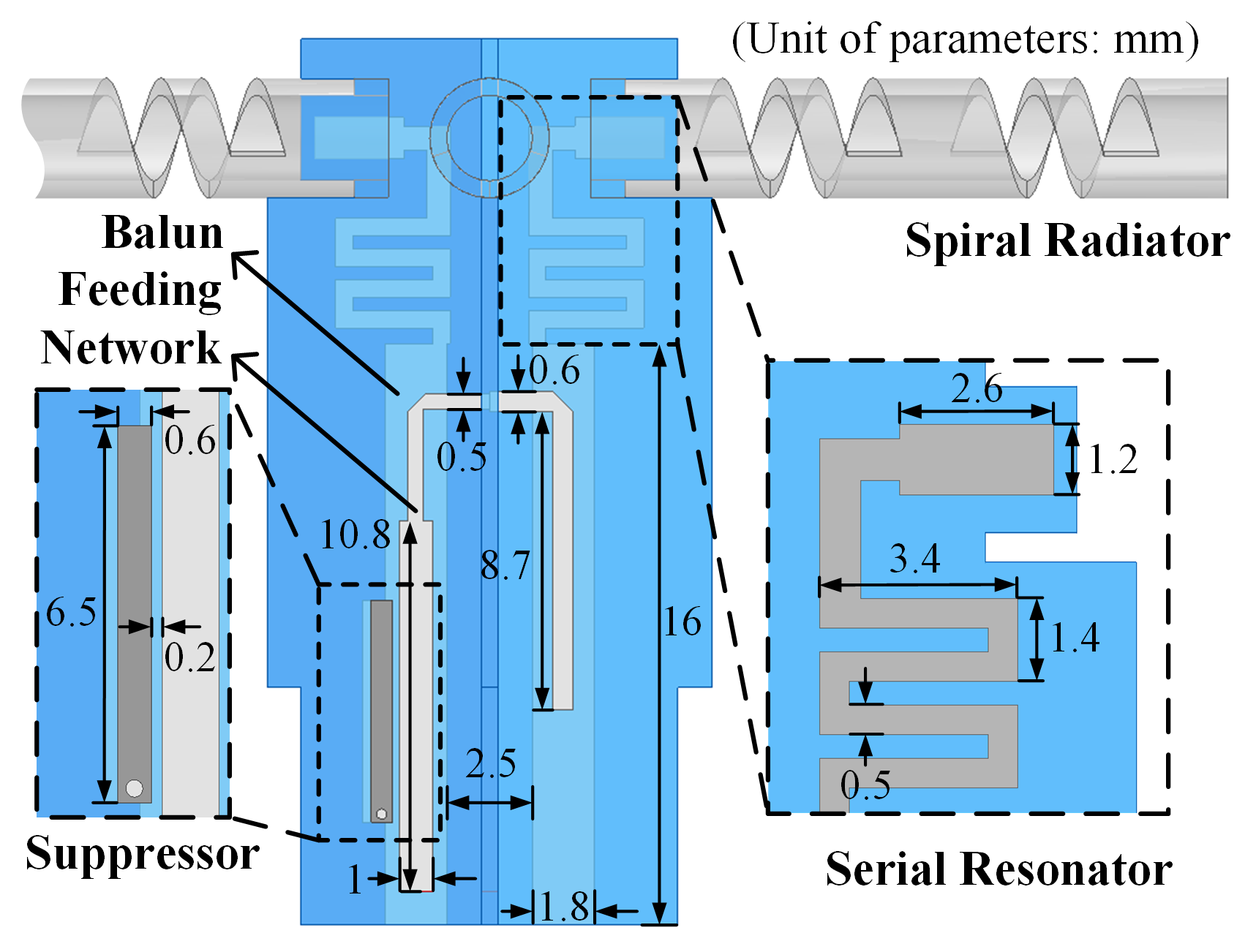}
\caption{Geometry of the LB segmented spiral antenna with integrated serial resonators and suppressors in the balun.}
\label{Balun}
\end{figure}

As shown in Fig. \ref{RCS}(b), the segmented spiral exhibits greater input impedance variation than the reference tube, making impedance matching more difficult. To restore the matching, as shown in Fig. \ref{Balun}, a pair of serial resonators is introduced between the LB radiators and the balun. The resonators, balun, and feeding network are printed on both sides of a Rogers 5880 substrate. A metal pad is inserted into the spiral radiator, forming a capacitor with the radiator’s metal structure, while a meander line between the capacitor and the balun acts as an inductor. Additionally, a suppressor composed of a strip, a via, and the ground is placed near the balun to improve the isolation between the LB and MB antennas in the MB.

\begin{figure}[!t]
\centering
\includegraphics[trim=0 10 0 0,scale=1]{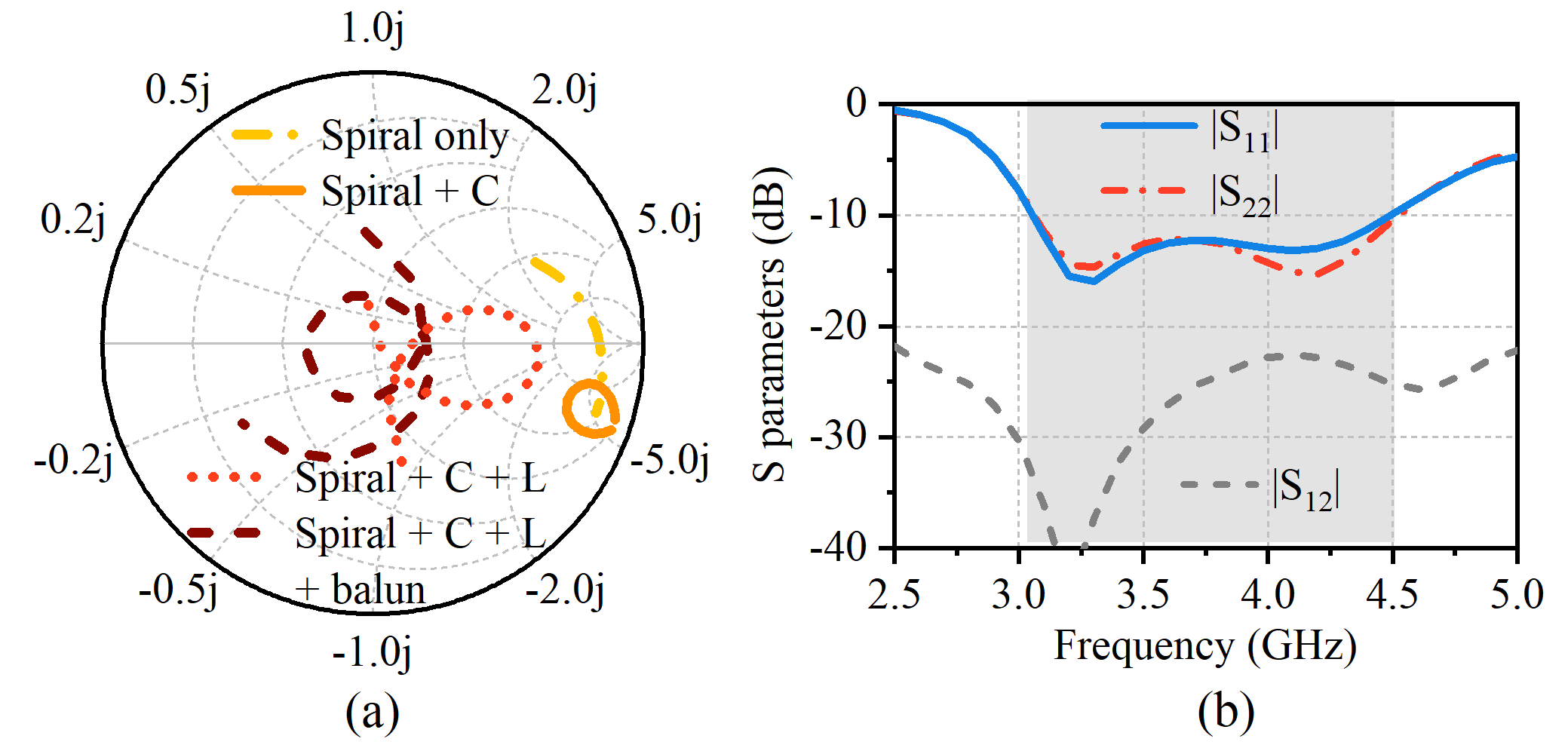}
\caption{(a) Input impedances of the LB segmented spiral dipole with different loading conditions. (b) Reflection coefficients ($|$S$_{11}$$|$ and $|$S$_{22}$$|$), and transmission coefficient $|$S$_{12}$$|$ of the LB segmented spiral antenna.}
\label{Smith}
\end{figure}

Fig. \ref{Smith}(a) shows the input impedances of the segmented spiral dipole under different conditions on the Smith chart, illustrating how the serial resonators restore impedance matching. The curves in Smith chart cover the frequency range of 3.0–4.9 GHz. The “Spiral only” curve lies far from the matching point with a large arc radius, making wideband matching difficult using a conventional balun. The addition of a capacitor decreases the arc radius and causes a counterclockwise rotation along the constant-resistance circle, resulting in the “Spiral + C” curve. As the metal pad area decreases, the rotation angle increases while the arc radius decreases. Introducing the inductor causes the curve to rotate clockwise, and as the meander line length increases, the “Spiral + C + L” curve moves closer to the matching point, indicating improved matching potential. Finally, incorporating the balun and feeding network enables the “Spiral + C + L + balun” curve to tightly encircle the matching point.

The S parameters of the matched LB antenna are presented in Fig. \ref{Smith}(b). The reflection coefficients are less than -10 dB in 3.05-4.51 GHz, and the polarization isolation exceeds 20 dB within this range. The wideband matching validates the effectiveness of the developed serial resonator.

\begin{figure}[!t]
\centering
\includegraphics[trim=0 10 0 5,scale=1]{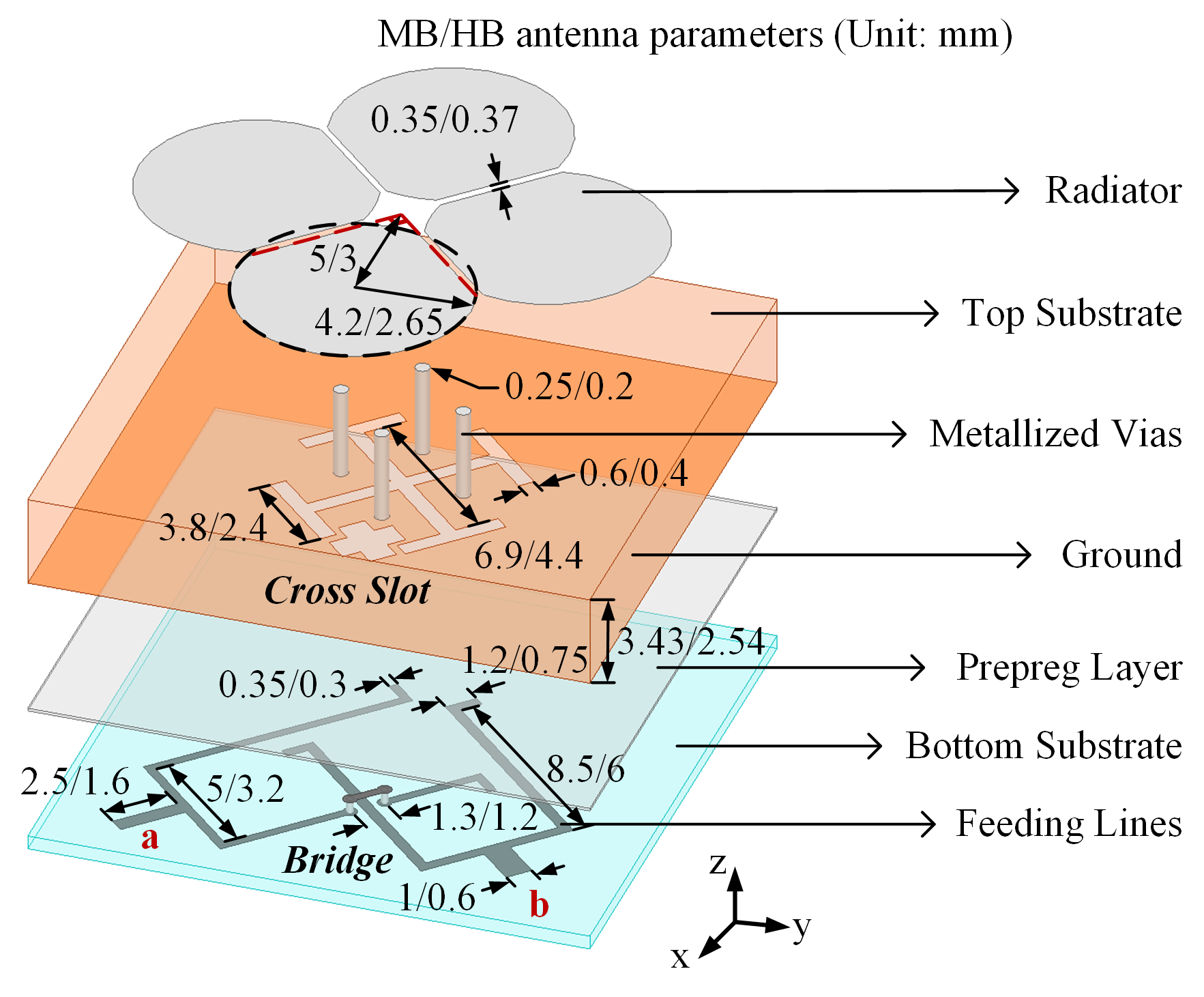}
\caption{Geometry of the MB and HB planar ME dipole antenna.}
\label{MB_HB}
\end{figure}

The designed planar magnetoelectric (ME) antenna, serving as the MB/HB element in the tri-band array, is shown in Fig. \ref{MB_HB}. The four-leaf-clover-shaped radiator is connected to the ground through four vias, while fork-shaped microstrip lines excite the radiator through a Jerusalem-cross-shaped slot in the ground. The proposed MB and HB planar ME dipoles operate over 6.3–10.0 GHz and 10.0–15.45 GHz, respectively, fully covering the 5G-Advanced and anticipated 6G band of 6.425–15.35 GHz.

In a tri-band antenna array, employing planar ME dipoles as the MB and HB elements helps prevent common-mode resonance \cite{S2} within the LB and MB, owing to their much lower profile compared to conventional dipoles. Moreover, the MB and HB radiators are positioned at nearly the same height, effectively mitigating HB scattering from the MB radiators.

\section{Holistic Suppression of Scattering and Coupling in The Tri-Band Array}

\begin{figure}[!t]
\centering
\includegraphics[trim=0 10 0 0,scale=1]{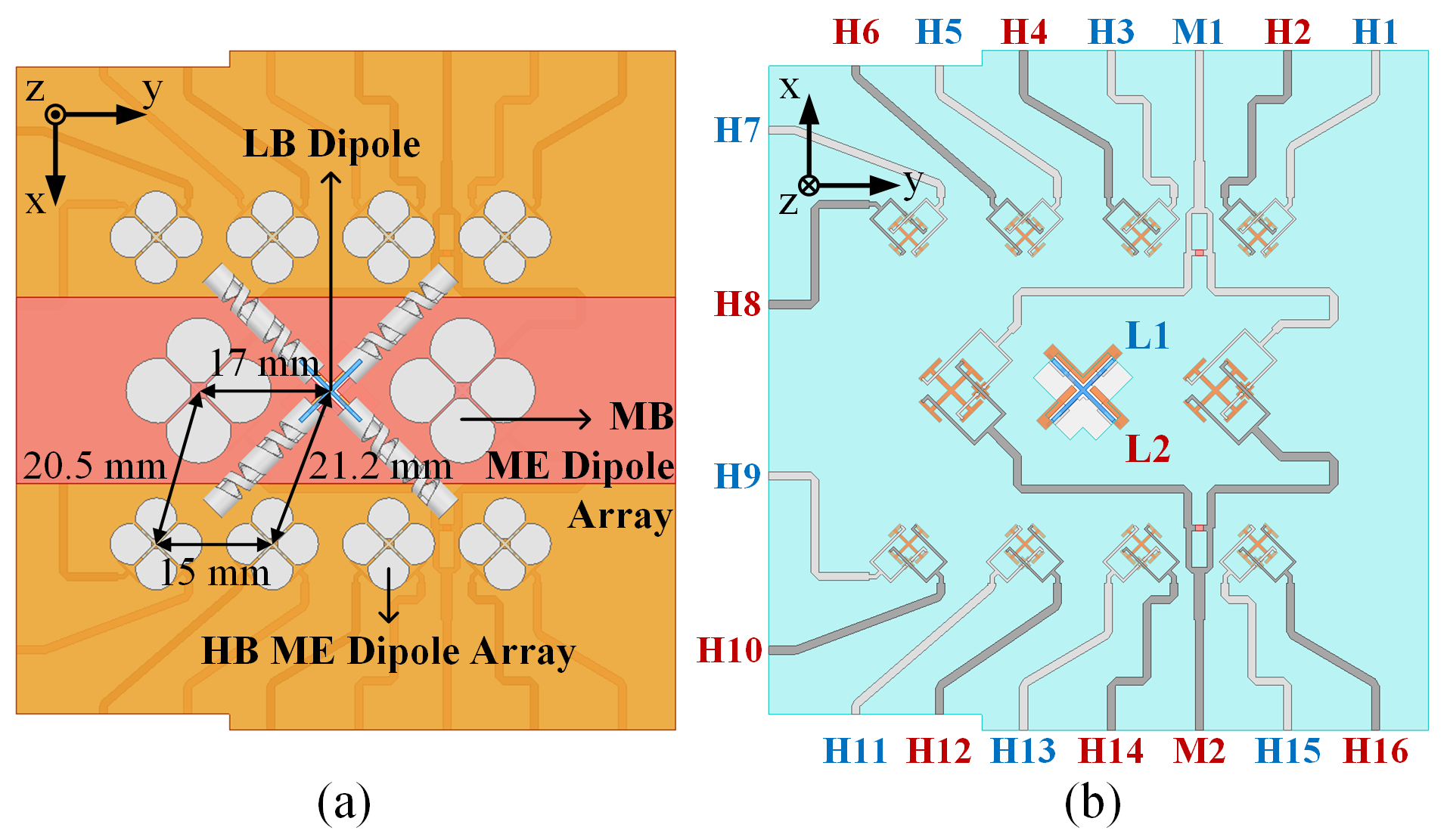}
\caption{(a) Top view and (b) bottom view of the tri-band
antenna array.}
\label{Tri_Band_Array_All_A}
\end{figure}

As shown in Fig. \ref{Tri_Band_Array_All_A}, the proposed LB, MB, and HB antennas are compactly arranged in an interleaved tri-band array to verify the effectiveness of the developed suppression methods. Although the MB and HB antennas share the same bottom substrate, their top substrate thicknesses differ. Consequently, the MB and HB arrays are designed as a planar structure consisting of three laminated dielectric layers.

\begin{figure}[!t]
\centering
\includegraphics[trim=0 10 0 0,scale=1]{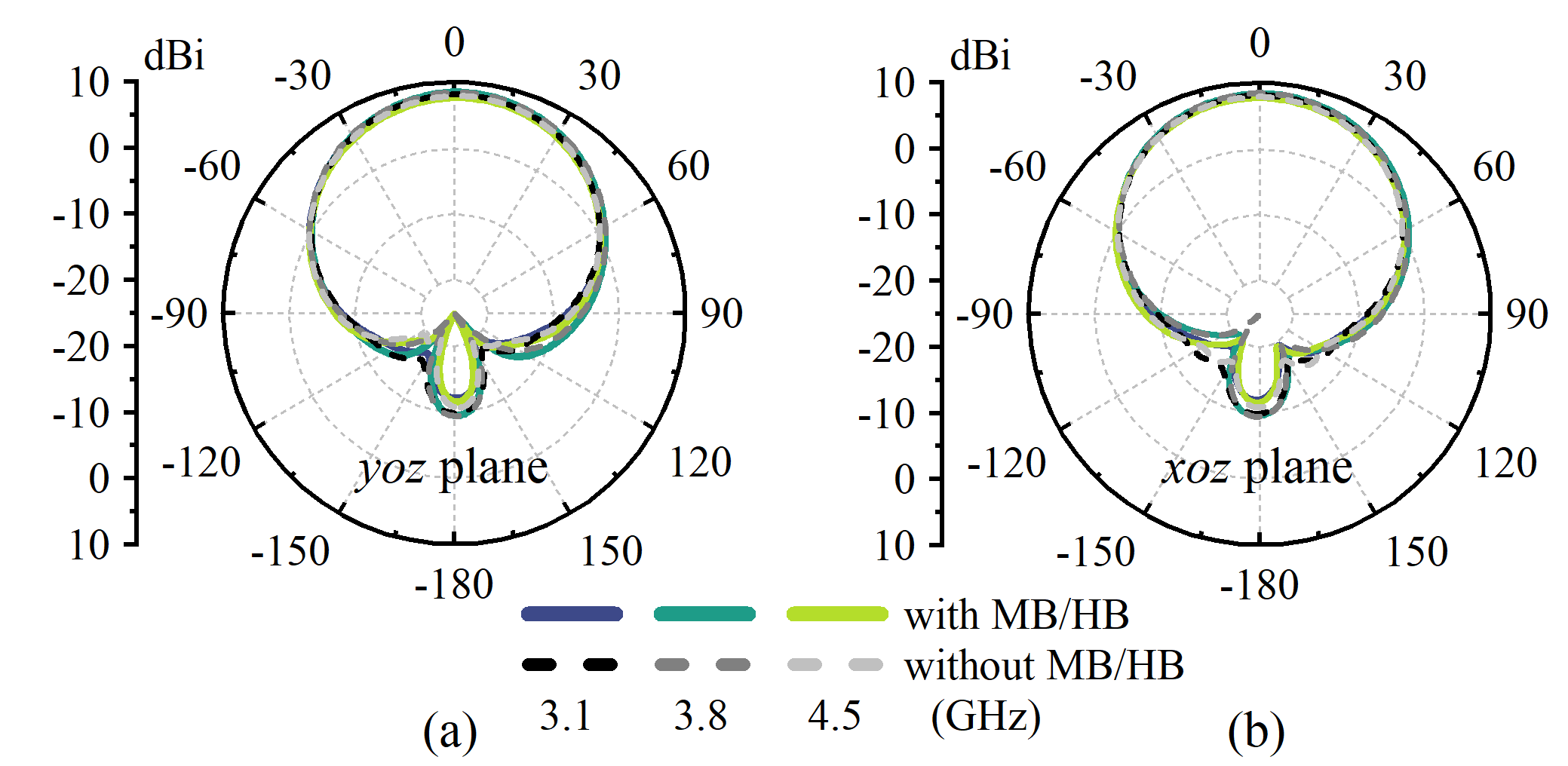}
\caption{Simulated radiation patterns of the LB antenna in (a) \textit{yoz} plane and (b) \textit{xoz} plane, when port L1 is excited.}
\label{Array_LB_Patterns_Ref}
\end{figure}

\begin{figure}[!t]
\centering
\includegraphics[trim=0 10 0 0,scale=1]{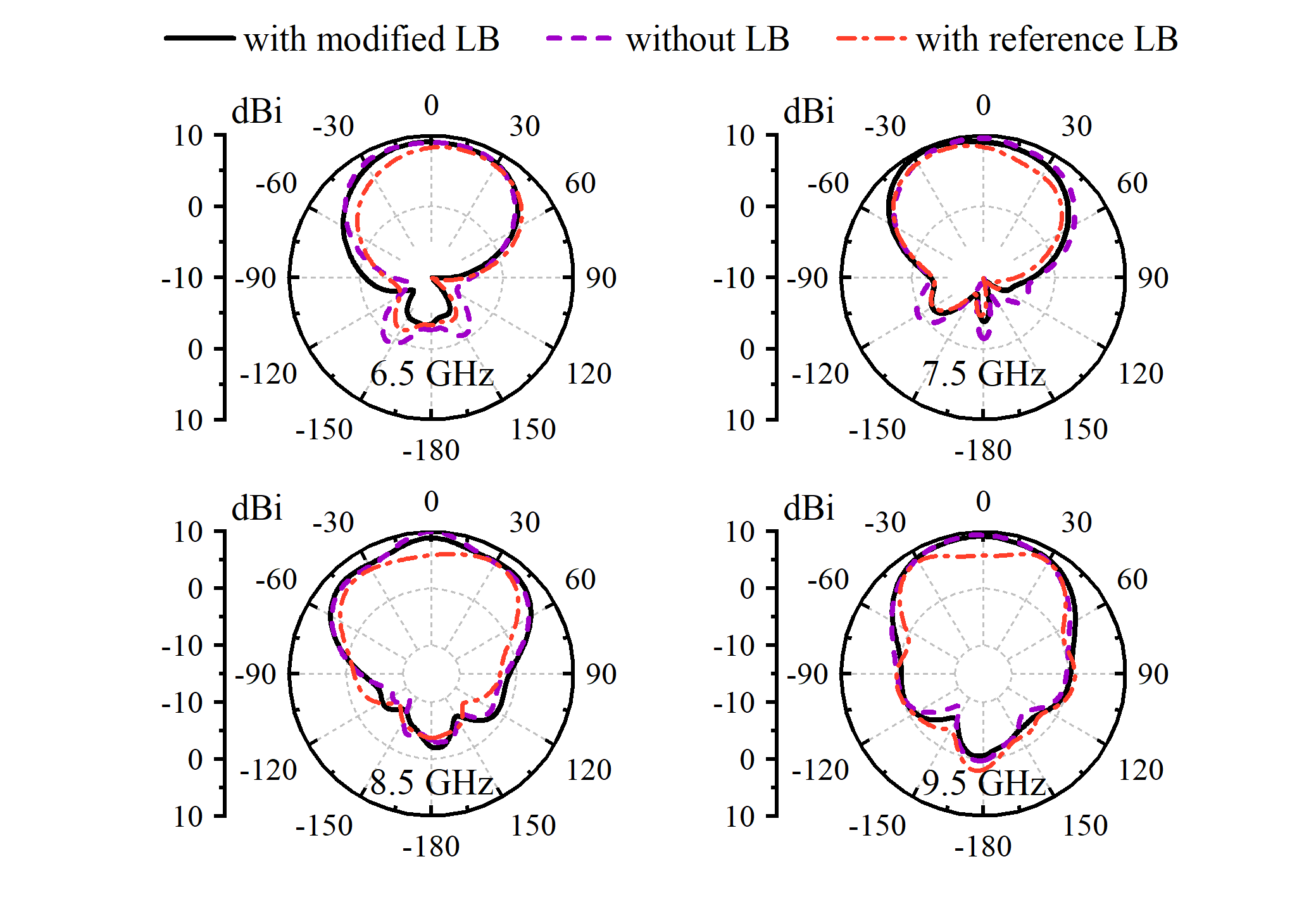}
\caption{Simulated radiation patterns of the MB antennas with and without the presence of different LB antennas, when port M1 is excited.}
\label{Array_MB_Patterns}
\end{figure}

\begin{figure}[!t]
\centering
\includegraphics[trim=0 10 0 0,scale=1]{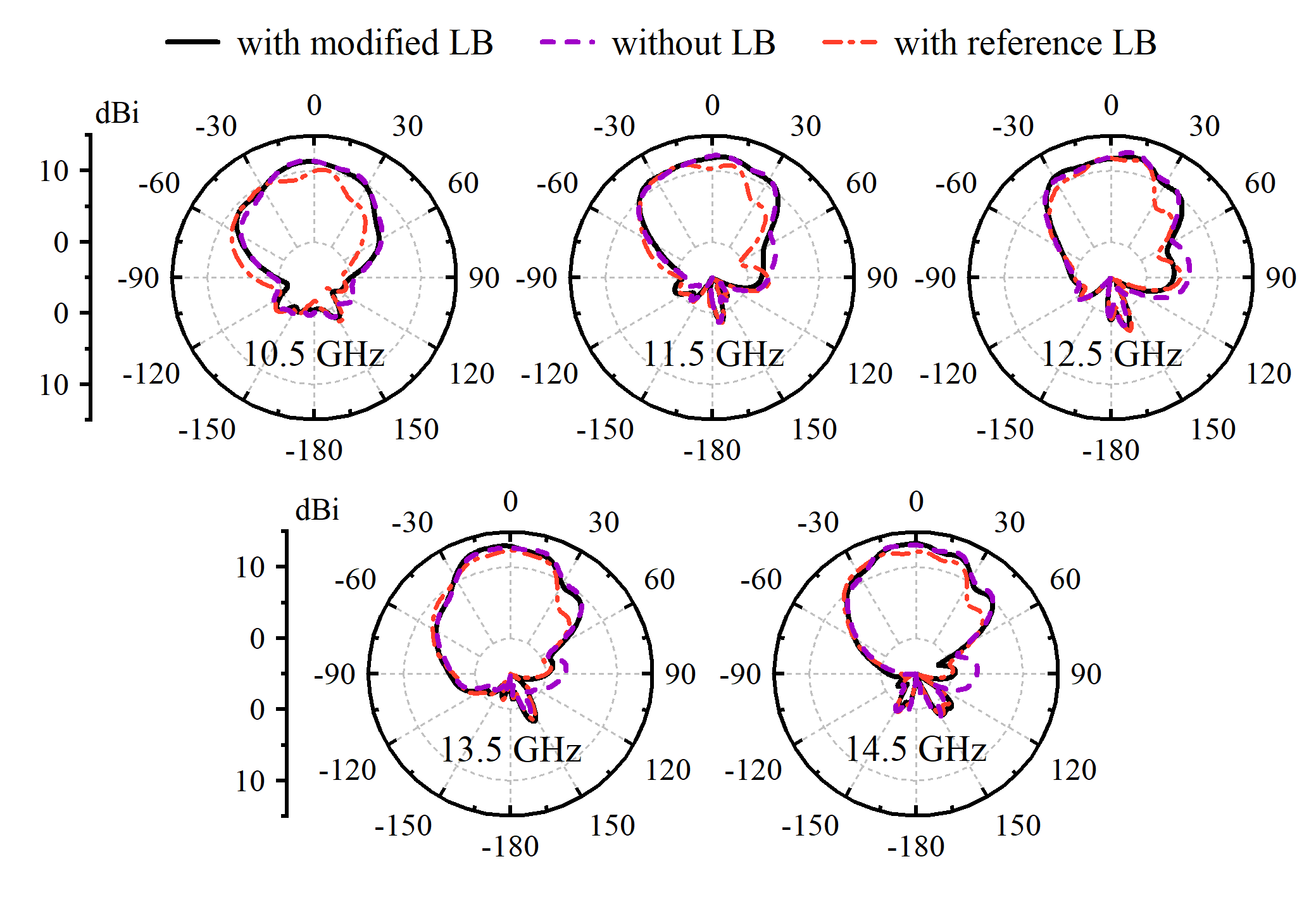}
\caption{Simulated radiation patterns of the HB antennas with and without the presence of different LB antennas, when ports H1-4 are excited.}
\label{Array_HB_Patterns}
\end{figure}

Fig. \ref{Array_LB_Patterns_Ref} compares the radiation patterns of the LB antenna with and without the MB and HB antenna array. Owing to the proposed planar ME dipole structure, the common-mode resonance induced by the MB and HB antennas is avoided within the LB. As a result, the LB antenna maintains undistorted radiation patterns across the entire band, closely matching those of the standalone LB antenna.

The MB and HB radiation patterns in the \textit{xoz} plane for different cases are shown in Fig. \ref{Array_MB_Patterns} and Fig. \ref{Array_HB_Patterns}, respectively. In the reference case, the conventional tube-based LB antenna severely distorts the MB/HB radiation patterns and reduces the realized gain. In contrast, adopting the LB antenna with the MB/HB-transparent segmented spiral minimizes these adverse effects, achieving MB/HB radiation patterns and realized gains consistent with those of the MB/HB antennas operating without the LB antenna.

\section{Experimental Results}

\begin{figure}[!t]
\centering
\includegraphics[trim=0 10 0 0,scale=1.1]{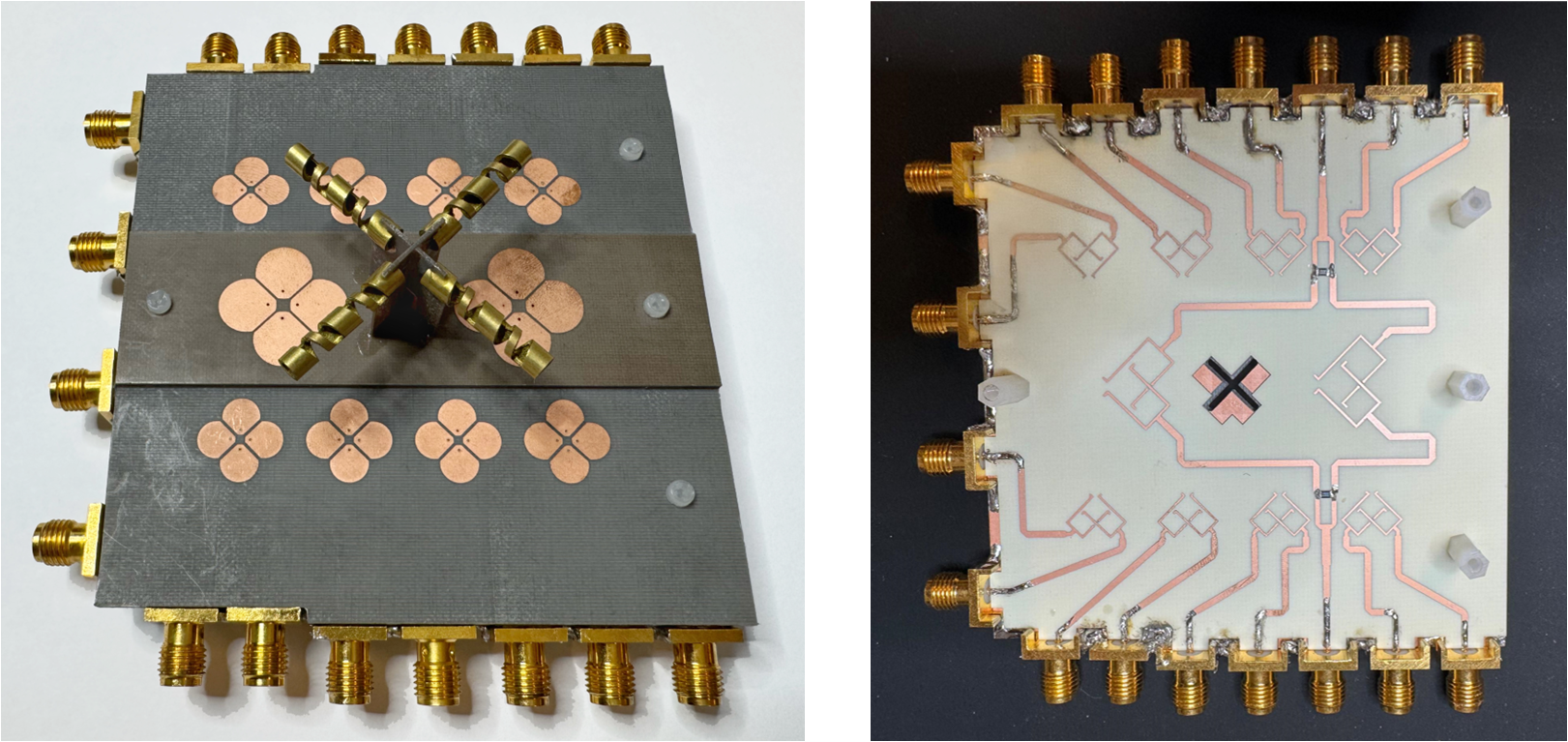}
\caption{Prototype of the tri-band antenna array.}
\label{Tri_Band_Array_All_B}
\vspace{-8pt}
\end{figure}

\begin{figure}[!t]
\centering
\includegraphics[trim=0 10 0 0,scale=1]{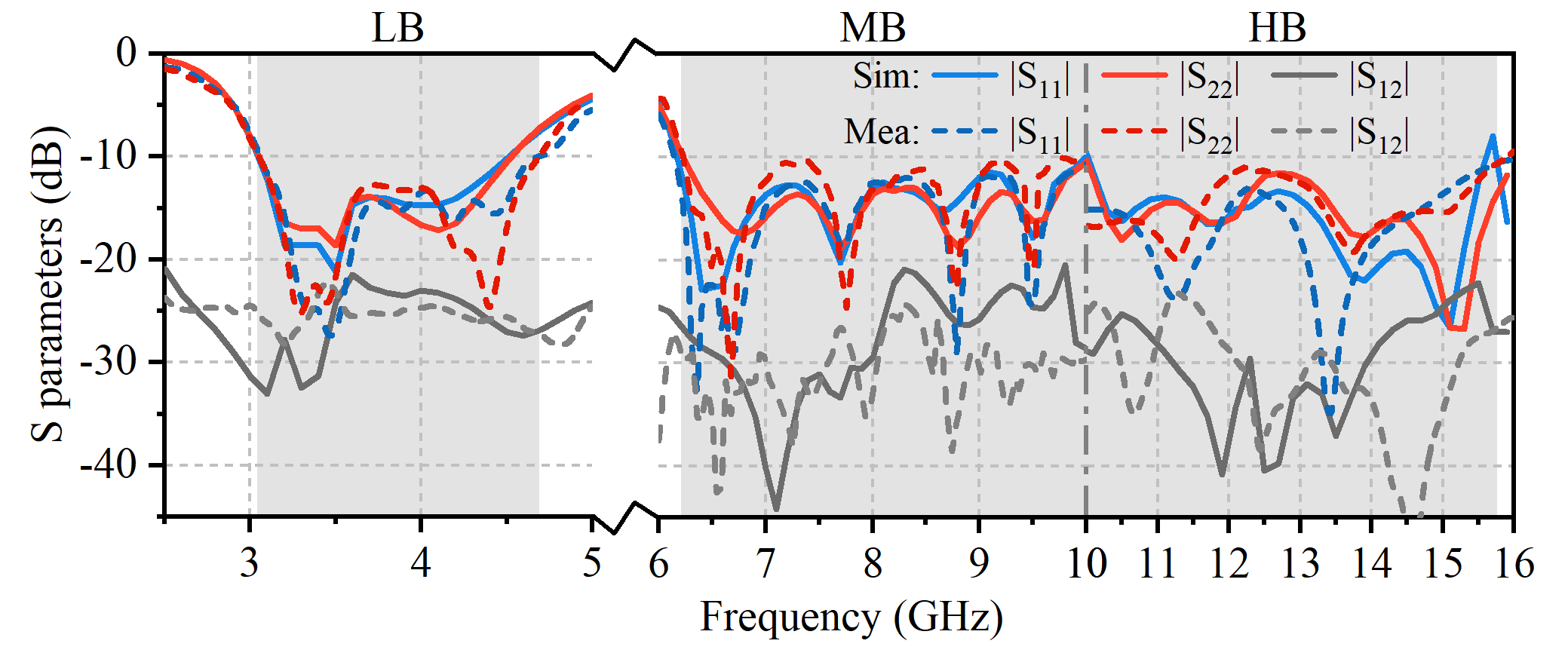}
\caption{Simulated and measured S-parameters of the LB, MB, and HB antennas in the tri-band array.}
\label{Array_LB_MB_HB_S}
\end{figure}

The prototype of the developed tri-band antenna array is shown in Fig. \ref{Tri_Band_Array_All_B}. The simulated and measured S-parameters are compared in Fig. \ref{Array_LB_MB_HB_S}. The measured reflection coefficients for the LB, MB, and HB antennas are below -10 dB over 3.05–4.68 GHz (42.2\%), 6.2–10.0 GHz (46.9\%), and 10.0–15.6 GHz (43.8\%), respectively, showing good agreement with the simulations.

\begin{figure}[!t]
\centering
\includegraphics[trim=0 10 0 5,scale=1]{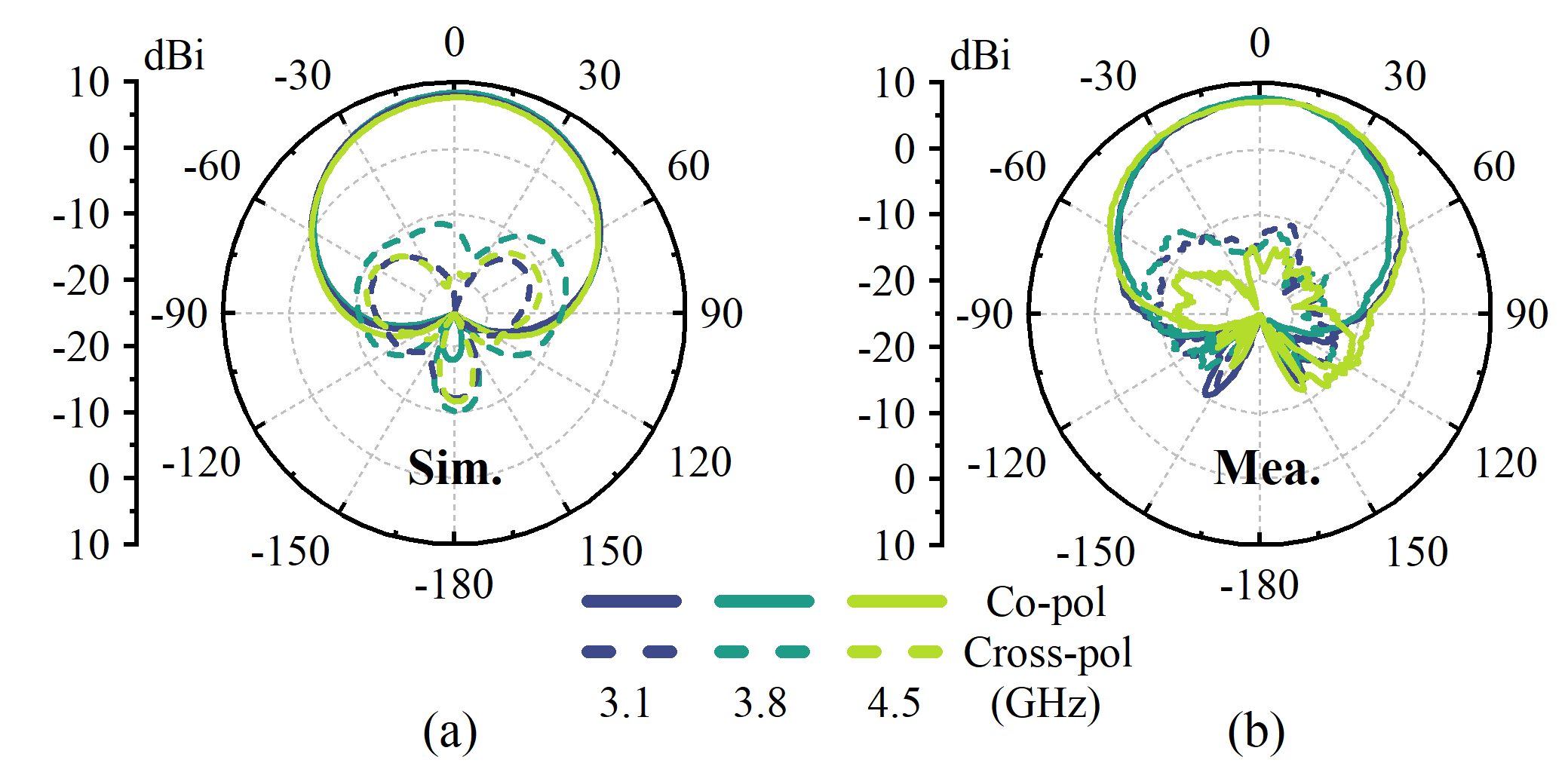}
\caption{(a) Simulated and (b) measured radiation patterns of the LB antenna in the \textit{yoz} plane, when port L1 is excited.}
\label{Array_LB_Patterns}
\end{figure}

\begin{figure}[!t]
\centering
\includegraphics[trim=0 15 0 5,scale=1]{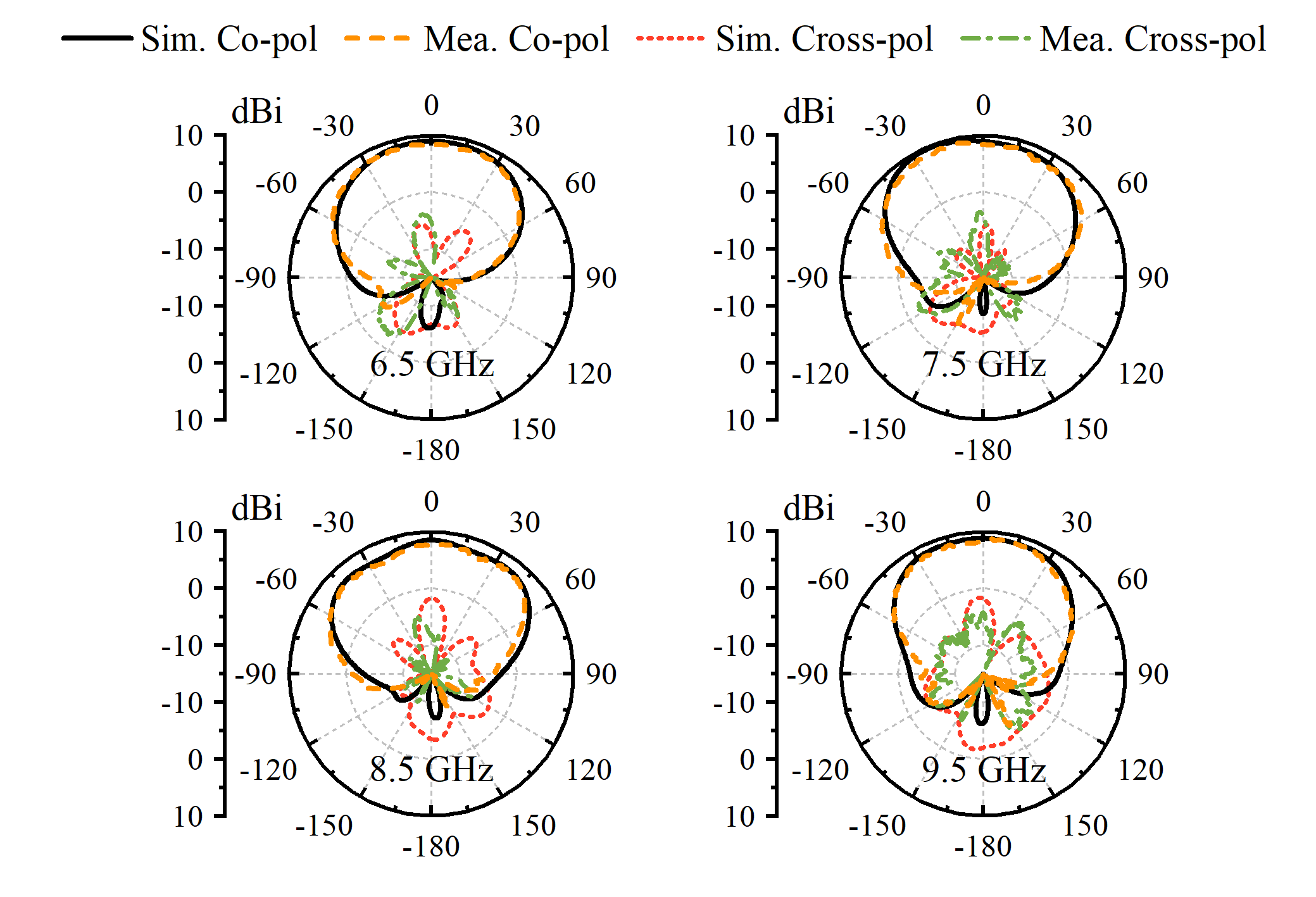}
\caption{Radiation patterns of the MB antennas in the \textit{xoz} plane.}
\label{Mea_Array_MB_Patterns}
\end{figure}

\begin{figure}[!t]
\centering
\includegraphics[trim=0 15 0 5,scale=1]{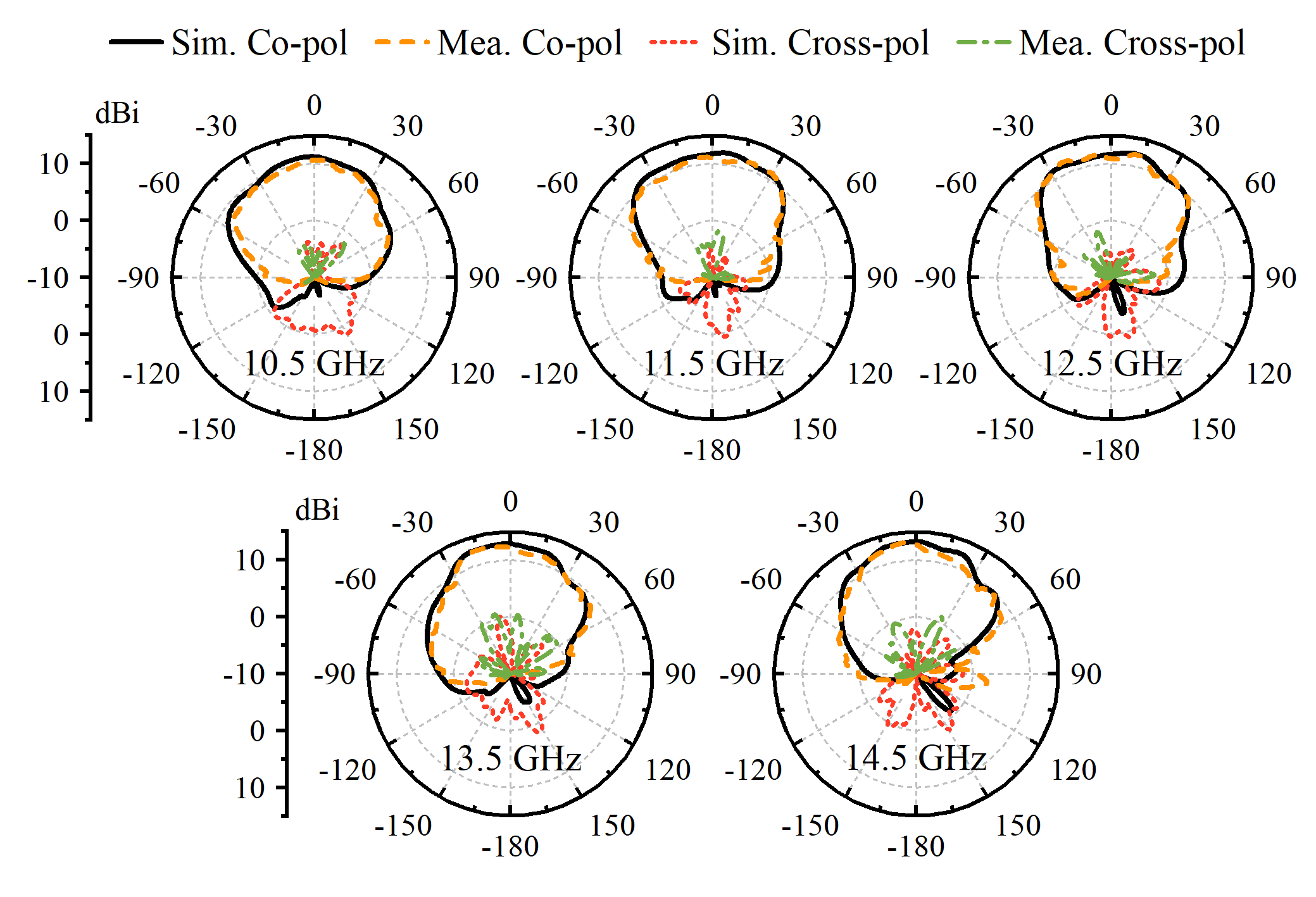}
\caption{Radiation patterns of the HB antennas in the \textit{xoz} plane.}
\label{Mea_Array_HB_Patterns}
\end{figure}

The simulated and measured radiation patterns of the LB, MB, and HB antennas in the tri-band array are compared in Figs. \ref{Array_LB_Patterns}, \ref{Mea_Array_MB_Patterns}, and \ref{Mea_Array_HB_Patterns}, respectively. The measured patterns closely agree with the simulated results, validating the effectiveness of the developed suppression methods.

\begin{figure}[!t]
\centering
\includegraphics[trim=0 10 0 0,scale=1]{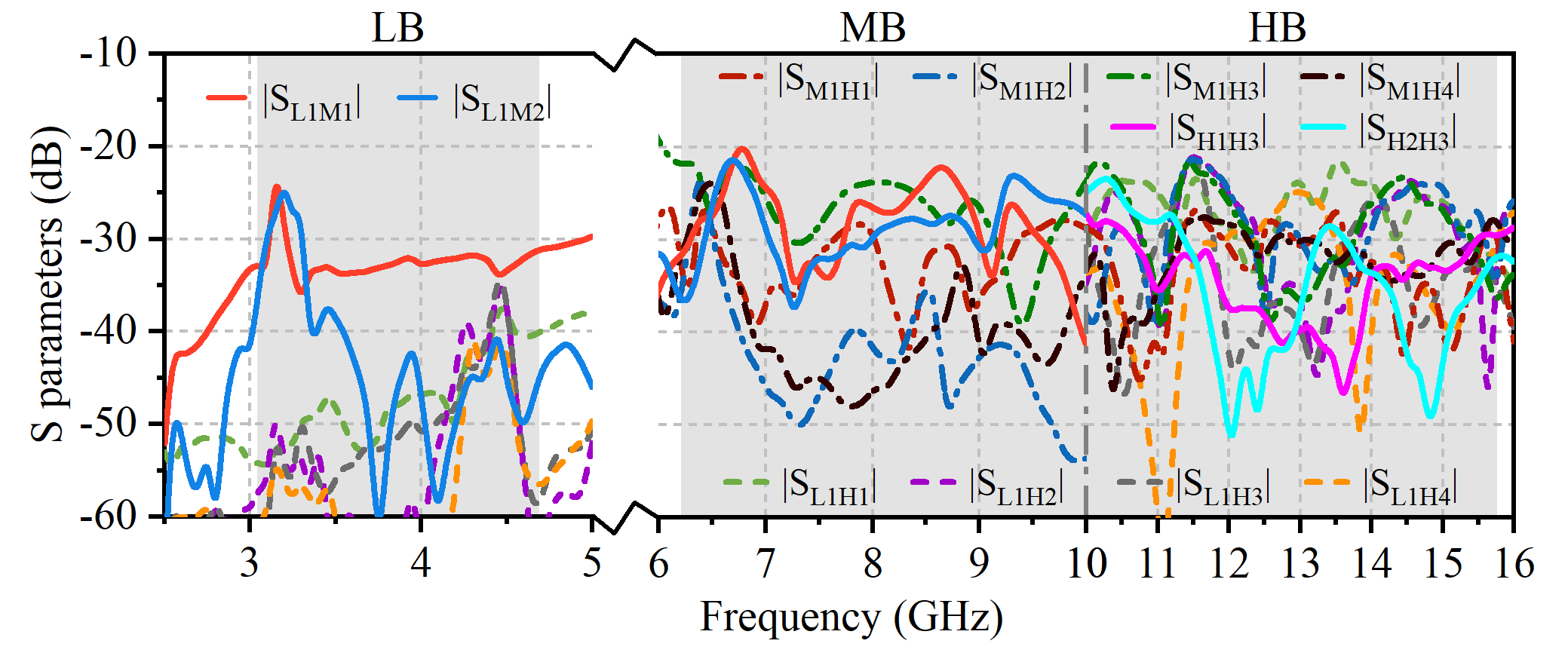}
\caption{Measured transmission coefficients between the LB, MB, and HB antennas in the tri-band antenna array (port numbers are labeled in Fig. 8).}
\label{Array_LB_MB_HB_iso}
\end{figure}

The measured isolation between any two ports in the proposed tri-band array exceeds 20 dB across all three bands, and representative results are given in Fig. \ref{Array_LB_MB_HB_iso}. The segmented spiral significantly reduces the MB- and HB-induced currents excited on the LB antenna, while the planar ME dipoles avoid the generation of resonant currents in the LB and MB. Thus, effective suppression of cross-band and in-band coupling is achieved over the three wide bands.

To demonstrate the advantages of the proposed techniques in this work, the performance is compared with other approaches, as listed in Table I. The proposed methods achieve suppression of scattering and coupling without increasing complexity of the array configuration, featuring the widest reported bandwidths across all three bands: LB, MB, and HB.

\begin{table}
 \centering
 \caption{Comparison of Techniques for Scattering and Coupling Suppression in Tri-Band Antenna Arrays}
 \label{table1}
 \vspace{-3pt}

 \begin{tabular}{ccc}
\toprule[1.2pt]
\specialrule{0em}{0.5pt}{0.5pt}
\toprule[1.2pt]
\textbf{\begin{tabular}[c]{@{}c@{}}Proposed structures \& \\Array dimensions\end{tabular}} & 
\textbf{\begin{tabular}[c]{@{}c@{}}Number of elements \& \\Operating band (GHz)\end{tabular}} & 
\textbf{\begin{tabular}[c]{@{}c@{}}Isolation\\(dB)\end{tabular}} \\ 
\midrule [1.2pt]

\begin{tabular}[c]{@{}c@{}}\cite{T5} Dual-band FSS radiator\\ + DRA \\0.71$\lambda_{L}$ $ \times $ 0.45$\lambda_{L}$ $ \times $ 0.16$\lambda_{L}$  \end{tabular} &
\begin{tabular}[c]{@{}c@{}}1 LB: 1.85--2.15 (15\%)\\ 2 MB: 3.4--3.6 (5.7\%)\\ 6 HB: 5.4--5.6 (3.6\%)\end{tabular} &
N/A \\ \midrule [1pt]

\begin{tabular}[c]{@{}c@{}}\cite{T1} Choke + Slot + Metal strip\\0.91$\lambda_{L}$ $ \times $ 0.84$\lambda_{L}$ $ \times $ 0.23$\lambda_{L}$  \end{tabular} &
\begin{tabular}[c]{@{}c@{}}1 LB: 0.79--0.96 (19.4\%)\\ 4 MB: 1.71--2.17 (23.7\%)\\ 32 HB: 3.4--3.6 (5.7\%)\end{tabular} &
N/A \\ \midrule [1pt]

\begin{tabular}[c]{@{}c@{}}\cite{T3} 45\degree{} rotation + slot \\+ FSS radiator\\0.65$\lambda_{L}$ $ \times $ 0.63$\lambda_{L}$ $ \times $ 0.2$\lambda_{L}$  \end{tabular} &
\begin{tabular}[c]{@{}c@{}}1 LB: 0.76--0.88 (14.6\%)\\ 6 MB: 1.9--2.7 (34.8\%)\\ 24 HB: 3.3--3.9 (16.7\%)\end{tabular} &
N/A \\ \midrule [1pt]

\begin{tabular}[c]{@{}c@{}}\cite{T2} Stacked arrangement\\ + FSS radiator + Helical cable\\0.69$\lambda_{L}$ $ \times $ 0.69$\lambda_{L}$ $ \times $ 0.177$\lambda_{L}$   \end{tabular} &
\begin{tabular}[c]{@{}c@{}}1 LB: 0.69--0.96 (32.7\%)\\ 4 MB: 1.8--2.7 (40.0\%)\\ 16 HB: 3.3--3.8 (14.1\%)\end{tabular} &
\textgreater{} 19 \\ \midrule [1pt]

\begin{tabular}[c]{@{}c@{}}Segmented spiral + Suppressor\\  + Planar ME dipole \\0.89$\lambda_{L}$ $ \times $ 0.86$\lambda_{L}$ $ \times $ 0.25$\lambda_{L}$ \end{tabular} &
\begin{tabular}[c]{@{}c@{}}1 LB: 3.05--4.68 (\textbf{42.2\%})\\ 2 MB: 6.2--10.0 (\textbf{46.9\%})\\ 8 HB: 10.0--15.6 (\textbf{43.8\%})\end{tabular} &
\textbf{\textgreater{} 20} \\
\bottomrule [1.2pt]
\specialrule{0em}{0.5pt}{0.5pt}
\bottomrule [1.2pt]
\end{tabular}
\vspace{-10pt}
\end{table}

\section{Conclusion}
This work proposes a tri-band shared-aperture 5G/6G antenna array featuring wideband scattering and coupling suppression. The suppression is achieved through a developed LB antenna employing segmented spirals, serial resonators, and suppressors, together with MB and HB antennas based on planar ME dipole structures. The LB (3.05–4.68 GHz, 42.2\%) antenna covers the 5G band of 3.3–4.2 GHz, while the MB (6.2–10.0 GHz, 46.9\%) and HB (10.0–15.6 GHz, 43.8\%) antennas collectively span the anticipated 5G-Advanced and 6G spectrum of 6.425–15.35 GHz. Over the three bands, the antennas maintain undistorted radiation patterns, and the isolation between any two ports in the antenna array exceeds 20 dB. The measured results are in good agreement with the simulations, confirming effective wideband suppression of scattering and coupling. Therefore, the proposed tri-band shared-aperture antenna array represents a promising candidate for future 6G applications.

% \IEEEtriggeratref{7}
% \IEEEtriggercmd{\enlargethispage{-20cm}}


%\bibliographystyle{IEEEtran}

\bibliographystyle{IEEEtran}
\bibliography{Reference}

\end{document}